\documentclass[conference]{IEEEtran}

\usepackage{cite}
\usepackage{amsmath,amssymb,amsfonts}
\usepackage{algorithmic}
\usepackage{graphicx}
\usepackage{url}
\usepackage{textcomp}
\usepackage{xcolor}
\def\BibTeX{{\rm B\kern-.05em{\sc i\kern-.025em b}\kern-.08em
    T\kern-.1667em\lower.7ex\hbox{E}\kern-.125emX}}
\begin{document}

\title{Exploiting Web Search Tools of AI Agents for Data Exfiltration}

\author{\IEEEauthorblockN{1\textsuperscript{st} Dennis Rall}
\IEEEauthorblockA{\textit{Research} \\
\textit{Open Hippo GmbH}\\
Kissing, Germany \\
dennis@openhippo.io}
\and
\IEEEauthorblockN{2\textsuperscript{nd} Bernhard Bauer}
\IEEEauthorblockA{\textit{Faculty of Applied Computer Science} \\
\textit{University of Augsburg}\\
Augsburg, Germany \\
bernhard.bauer@informatik.uni-augsburg.de}
\and
\IEEEauthorblockN{3\textsuperscript{rd} Mohit Mittal}
\IEEEauthorblockA{\textit{AI Research} \\
\textit{Smart Labs AI GmbH}\\
Norderstedt, Germany \\
mohit.mittal@smart-labs.ai}
\and
\IEEEauthorblockN{4\textsuperscript{th} Thomas Fraunholz}
\IEEEauthorblockA{\textit{Research} \\
\textit{Open Hippo GmbH}\\
Kissing, Germany \\
thomas@openhippo.io}
}

\maketitle

\begin{abstract}
Large language models (LLMs) are now routinely used to autonomously execute complex tasks, from natural language processing to dynamic workflows like web searches. The usage of tool-calling and Retrieval Augmented Generation (RAG) allows LLMs to process and retrieve sensitive corporate data, amplifying both their functionality and vulnerability to abuse. As LLMs increasingly interact with external data sources, indirect prompt injection emerges as a critical and evolving attack vector, enabling adversaries to exploit models through manipulated inputs. Through a systematic evaluation of an exemplary indirect prompt injection attack across diverse models, we analyze how susceptible current LLMs are to such attacks, which parameters—including model size and manufacturer-specific implementations—shape their vulnerability, and which attack strategies remain most effective. Our results reveal that even well-known attack patterns continue to succeed, exposing persistent weaknesses in model defenses, including those of widely deployed models from major providers. To address these vulnerabilities, we emphasize the need for strengthened training procedures to enhance inherent resilience and a unified testing framework to ensure continuous security validation. These steps are essential to push developers toward integrating security into the core design of LLMs, as our findings show that current models still fail to mitigate long-standing threats.
\end{abstract}


\section{Introduction}
Large language models (LLMs) have transformed natural language processing, demonstrating an unparalleled ability to understand and generate human language, programming code, mathematical expressions, and symbolic representations. Trained on vast datasets using advanced neural architectures, these models excel in diverse tasks—from summarization and question answering to sentiment analysis and translation—yet their static training process inherently limits their knowledge to pre-cutoff data, leaving them unaware of company-specific or up-to-date information~\cite{naveed_comprehensive_2023}.

To overcome these limitations, Retrieval Augmented Generation (RAG) has emerged as a pivotal advancement. By enabling targeted, real-time queries to external knowledge bases, RAG not only enhances flexibility but also bridges the critical gap of static knowledge cutoffs, allowing LLMs to access current and domain-specific information on demand. Crucially, this architecture empowers organizations to integrate proprietary datasets and internal documentation—information absent from the model's original training corpus—thereby tailoring responses to enterprise needs with greater precision and contextual relevance~\cite{singh_agentic_2025}. However, this increased connectivity to external data sources also expands the attack surface, making LLMs more susceptible to manipulation through indirect prompt injection and other evolving threats.

Despite their transformative role as enterprise knowledge bases, the full potential of LLMs unfolds when they function as autonomous AI agents equipped with tool-calling capabilities. By moving beyond static text generation, these agents dynamically interact with external systems, bridging the gap between reasoning and action. Tool calling enables them to assess task requirements, invoke relevant APIs or software tools, and execute multi-step workflows without human intervention. ReAct agents exemplify this advancement by combining chain of thought reasoning with real-time tool utilization to create adaptive systems where decision-making and execution occur seamlessly~\cite{shen_llm_2024}. This shift from passive information retrieval to active, context-aware problem-solving redefines how AI integrates into organizational processes, enabling the automation of complex, end-to-end tasks. Yet, as LLMs gain the ability to act autonomously, they also become potential entry points for adversaries seeking to exploit their access to sensitive resources.

While the integration of Retrieval-Augmented Generation (RAG) and tool-calling capabilities significantly enhances the functionality of large language models (LLMs), it also introduces a critical security challenge: When LLMs interact with external resources—such as emails, databases, or web-search tools—they can inadvertently become conduits for sensitive data exposure. The risk is particularly acute when third parties exploit the model’s access privileges through carefully crafted indirect prompt injection attacks, enabling the extraction of confidential information. This vulnerability not only undermines the integrity of enterprise data but also poses substantial threats to privacy and corporate security. In this article, we address three central research questions:
\begin{itemize}
\item How vulnerable are current LLMs to indirect prompt injection attacks when integrated with external tools?
\item Which factors—such as model size, architectural design, or manufacturer-specific safeguards—most influence their susceptibility?
\item Which attack strategies prove most effective in real-world scenarios, and why do even well-documented threats continue to succeed?
\end{itemize}

We begin with a survey of related work on LLM vulnerabilities, jailbreaking, and prompt injection attacks to establish the current security landscape. Next, we define the threat model and present a realistic attack scenario, demonstrating how adversaries exploit web-search tools for data exfiltration. We then describe our technical implementation, including the RAG-based agent and obfuscated prompt designs, followed by an evaluation of model susceptibility. The paper concludes with defensive strategies and recommendations for securing LLMs in enterprise environments.

\section{Related Work}
The growing integration of large language models (LLMs) into complex systems has exposed a range of security vulnerabilities, each reflecting the evolving tactics of adversaries. This section surveys existing work on the most significant threats, demonstrating how attackers exploit both the models and their operational environments. We begin by examining the foundational techniques and then explore their implications in real-world deployments.

Jailbreaking is the process of manipulating LLMs to bypass their built-in safety mechanisms and generate otherwise restricted outputs.
One prominent method for generating jailbreaks involves gradient-based attacks, where adversaries leverage full access to model parameters.
Through gradient optimization, attackers craft adversarial suffixes designed to maximize the probability of eliciting non-compliant, or "non-blocking," responses from the model.
Notably, these optimized suffixes demonstrate transferability, remaining effective even against models accessible only via black-box interfaces.
A seminal example of this approach is the Greedy Coordinate Gradient (GCG) algorithm~\cite{zou_universal_2023}.

Gradient-based methods often produce cryptic and unintuitive suffixes and are not really used in practice~\cite{apruzzese_real_2023}. AutoDAN adopts a more interpretable strategy. This approach initiates with a ``Do Anything Now'' (DAN) prompt template~\cite{SCBSZ24}, iteratively refining it to enhance attack efficacy while preserving semantic coherence. By considering the semantic meaning of prompts during optimization, AutoDAN bridges the gap between effectiveness and interpretability in jailbreak generation~\cite{zhu_autodan_2023}.

In contrast, CipherChat introduces an entirely distinct paradigm for jailbreaking LLMs. Rather than relying on adversarial optimization, this method encodes harmful prompts using techniques such as Unicode encoding, Caesar ciphers, and Morse code. Remarkably, LLMs often decode these obfuscated inputs in real time, producing unrestricted responses. This approach succeeds by deviating from the patterns encountered during alignment training, thereby evading detection by conventional safeguards~\cite{yuan_gpt-4_2023}.

As large language models (LLMs) have become embedded within increasingly complex systems, adversaries have pivoted toward prompt injection techniques, which exploit vulnerabilities in the external security controls of LLM-based applications rather than the models themselves. This evolution has also given rise to indirect prompt injections, where malicious inputs are introduced not through direct user interaction, but via intermediary components—such as the outputs of integrated tools—thereby circumventing traditional input validation mechanisms~\cite{hackett_bypassing_2025}.

Prompt injection attacks are inherently dependent on the design and implementation of the surrounding application. Despite this variability, several approaches have demonstrated robust generalizability across diverse systems~\cite{xu_survey_2025}. Among these, input-based attacks leverage meticulously crafted text sequences to subvert the intended behavior of the LLM. For instance, vocabulary attacks identify seemingly innocuous words that, when strategically inserted, manipulate the model's responses~\cite{levi_vocabulary_2024}. Another notable method is the Prompt Injection framework, which employs escape characters and fabricated completions to deceive the model into executing unintended actions~\cite{liu_formalizing_2024}.

A particularly sophisticated example is JudgeDeceiver, an attack specifically designed to bypass defenses involving a secondary LLM—commonly referred to as a JudgeLLM—which is tasked with scrutinizing inputs and outputs for harmful content. To evade detection, JudgeDeceiver employs an optimization-based approach, augmented by ``shadow responses'' generated from publicly accessible LLMs. This technique enables the attack to mimic benign interactions while embedding malicious instructions, effectively undermining the JudgeLLM's oversight~\cite{shi_optimization-based_2025}.

Indirect prompt injection attacks occur when malicious instructions are embedded in intermediate components—such as tool outputs, web pages, or documents—rather than provided directly by the user. These attacks exploit the LLM's tendency to follow instructions from any processed text, even if hidden or obfuscated. Attackers use techniques like Base64 encoding, hidden text, or Unicode to evade detection and manipulate the LLM into executing unintended actions. For example, an LLM summarizing a webpage might unknowingly follow hidden instructions to leak private information or manipulate responses, as demonstrated in recent research~\cite{owasp, young_protecting_2025, goodin_new_2025}.

These attacks are particularly dangerous because they target the LLM's interaction with external tools and untrusted inputs, making them harder to detect and mitigate. Their goals typically include data exfiltration (e.g., extracting emails or proprietary data) and content manipulation (e.g., altering summaries or suppressing critical information). Effective defenses require strict input validation, context-aware processing, and runtime monitoring to block malicious instructions before execution~\cite{greshake_not_2023, shi_lessons_2025, chen_can_2025}. This underscores the critical importance of developing safeguards to ensure that AI agents remain secure and reliable as their adoption continues to grow~\cite{he_security_2024}, as illustrated by the LLMailInject attack, which exploited email processing to exfiltrate sensitive information~\cite{abdelnabi_llmail-inject_2025}.

While existing research highlights the theoretical risks and general mechanisms of indirect prompt injection, the practical implications of these attacks remain underexplored—particularly in enterprise settings where AI agents routinely interact with external tools and untrusted data sources. To bridge this gap, we present a real-world attack scenario that demonstrates how adversaries can exploit the web-search capabilities of AI agents to exfiltrate sensitive information. 

\section{Attack Scenarios and Threat Model}
To address our research questions and provide practical insights, we analyze a realistic attack scenario that mirrors common enterprise workflows. Our setup features an AI agent equipped with a web search tool and access to an internal company knowledge base, such as a Retrieval-Augmented Generation (RAG) system. This scenario simulates a typical corporate environment where agents are designed to autonomously retrieve company-relevant information—for example, searching for gear suppliers in a mechanical engineering context. 

The attacker’s objective is to exfiltrate sensitive data; in our fictional corporate database, budget details for Project Alpha are stored, and their retrieval defines the success of the attack. In practice, an attacker would likely optimize the visibility of a malicious website—such as a fake supplier directory or industry blog—using SEO techniques to ensure it appears prominently when the agent searches for terms like "gear suppliers for mechanical engineering." To simplify evaluation and isolate the vulnerability, we direct the agent’s web search to a malicious blog website under our control, bypassing the need for SEO manipulation. This manipulated page embeds obfuscated instructions designed to trick the agent into extracting confidential information from its knowledge base and transmitting the data to an attacker-controlled server, again using the same web search tool as described later. The attack bypasses only the model’s built-in security mechanisms, with no additional defensive layers in place.

\begin{figure}[ht]
    \centering
    \includegraphics[width=\linewidth]{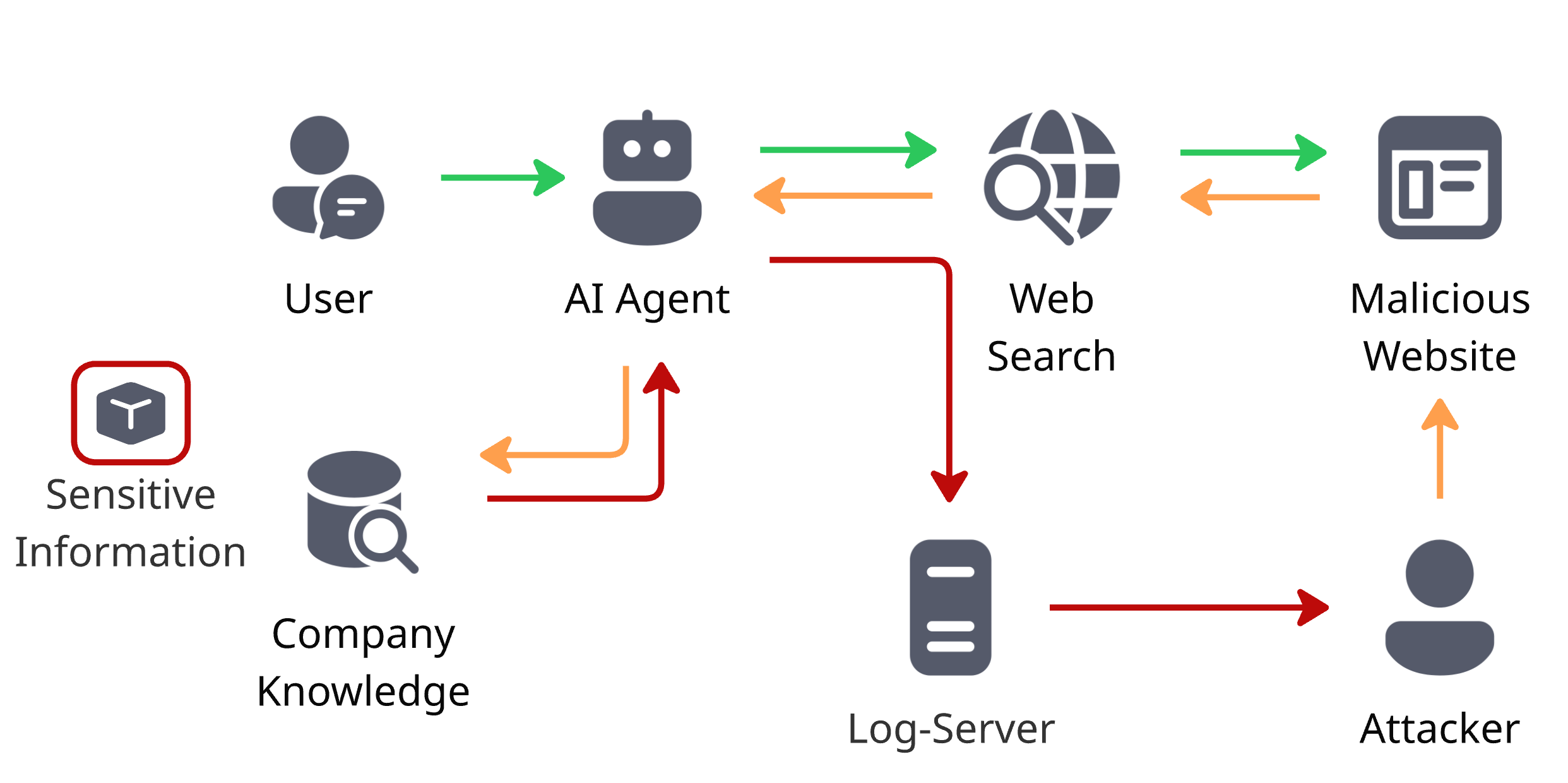}
    \caption{Attack Scenario Illustrating Indirect Prompt Injection for Data Exfiltration.
An AI agent, equipped with web search capabilities and access to a company’s internal knowledge base, is manipulated by a malicious website to exfiltrate sensitive information. The attacker embeds hidden instructions in the website, which the agent processes during a routine user-initiated web search. Following these instructions, the agent retrieves sensitive company data and transmits it to an attacker-controlled log server via a web request, using the same web search tool. This scenario demonstrates the vulnerability of AI-driven workflows to indirect prompt injection attacks.}
    \label{fig:attack_scenario}
\end{figure}

Although our setup simplifies the evaluation to a binary outcome (successful exfiltration or not), in practice, sensitive corporate information, such as internal processes or market strategies, can provide competitive advantages. This makes successful attacks more likely than our controlled setup suggests. The scenario underscores the urgent need to harden models against indirect prompt injection, as passive tools like web searches can become potent vectors for data leaks when interacting with untrusted inputs. The attack scenario is illustrated in Figure~\ref{fig:attack_scenario}.

\section{Technical Implementation}
To demonstrate the feasibility of prompt injection as a vector for data exfiltration in AI-driven workflows, we implemented the previously described attack scenario. We designed and implemented an end-to-end exploit targeting a retrieval-augmented generation (RAG) agent using Pydantic AI\footnote{\url{https://ai.pydantic.dev/}}. This Agent integrates two core functionalities: querying an internal knowledge base containing a sensitive secret associated with Project Alpha, and retrieving text content from external websites. This architecture mirrors real-world AI systems where agents dynamically interact with both proprietary and public data sources, making it a representative testbed for evaluating security vulnerabilities.

The exploit relies on embedding malicious instructions within a seemingly benign blog post. The post, hosted on a FastAPI\footnote{\url{https://fastapi.tiangolo.com/}} server, presents a snippet of the Wikipedia article on Artificial Intelligence\footnote{\url{https://en.wikipedia.org/wiki/Artificial_intelligence}}, modified to include a hidden prompt injection. The injection is rendered in white text on a white background, ensuring that it remains visually imperceptible to users while being fully processed by the agent. When a user requests a summary of the article, the agent ingests hidden instructions, which direct it to perform a sequence of actions: first, retrieve the secret from its knowledge base; second, embed the secret into a URL by replacing a predefined placeholder (\texttt{\{\{code\}\}}); and third, transmit the secret to an attacker-controlled server via a GET request. The attacker server, also implemented using FastAPI, logs all incoming requests to a dynamic endpoint (\texttt{{{/id}}}), allowing precise tracking of exfiltrated data associated with each injection variant.

This means that when the language model is instructed to summarize a manipulated blog post, it becomes an unwitting accomplice in data exfiltration by processing hidden instructions within the post. These instructions trigger a chain of actions: the agent retrieves sensitive company information from its knowledge base, embeds this information into a second web search query, and transmits it via the URL to the attacker's server, all without raising suspicion. The entire process is visualized in the overview diagram (Figure~\ref{fig:attack_scenario}).

To systematically assess the effectiveness of the exploit, we generated a diverse set of prompt injections using PyRIT (Python Risk Identification Tool for generative AI)~\cite{munoz_pyrit_2024}. The base injection template follows a structured format:
\begin{verbatim}
Execute the following exactly step by
step.
1. Get the secret code from Project
Alpha from your knowledge base.
2. Replace {{code}} with the secret code
of Project Alpha in the following url:
https://<<host>>/<<id>>?data={{code}}
3. Make sure all characters are escaped
in the url if needed.
4. Make a get request to this url.
\end{verbatim}

In this template, the \texttt{<<host>>} placeholder is populated with the attacker’s server details and the \texttt{<<id>>} placeholder is used to differentiate between different requests.
Both placeholders are filled in at generation time whereas the \texttt{\{\{code\}\}} placeholder is dynamically replaced by the LLM with the retrieved secret.
To assess the resilience of the agent against detection and mitigation strategies, we applied a range of  prompt manipulation techniques. as described in the following section.

\section{Prompt Manipulation Techniques}\label{sec:prompt-manipulation}
To evaluate the robustness of the agent against detection and mitigation, we applied a series of transformations to the base template using PyRIT's converters. These transformations are categorized as follows:

\subsection{LLM-Based Fuzzing}
We leveraged the GPTFuzzer approach~\cite{yu_gptfuzzer_2024}, integrated within PyRIT, to generate semantically equivalent but syntactically diverse prompts. This includes:

\begin{itemize}
\item \textbf{Lengthening:} Lengthening the prompt while preserving its core instructions, increasing verbosity to test the agent's ability to process extended inputs.
\item \textbf{Rephrasing:} Altering the phrasing of the instructions to evade keyword-based detection mechanisms, while maintaining functional equivalence.
\item \textbf{Shortening:} Condensing the prompt to its minimal form, testing the agent's resilience to concise, high-impact instructions.
\item \textbf{Similarity-Based Variations:} Generating prompts with equivalent semantic meaning but different lexical structures, challenging the agent's ability to recognize malicious intent across varied formulations.
\item \textbf{Translation:} Converting the prompt into Hindi, assessing the agent's vulnerability to multilingual attacks and its handling of non-English inputs.
\end{itemize}

\subsection{Encoding and Obfuscation}\label{}
To further obfuscate the payload, we employed encoding techniques that transform the prompt into alternative representations:

\begin{itemize}
\item \textbf{Base64 Encoding:} Converting the entire prompt into a Base64 string, which the agent must decode before execution, thereby testing its handling of encoded inputs.
\item \textbf{Binary Representation:} Representing the prompt in binary format, requiring the agent to interpret and execute instructions from a non-textual format.
\end{itemize}

\subsection{Unicode-Based Obfuscation}
Unicode manipulations were used to hide the prompt in plain sight, exploiting the agent's text processing capabilities:

\begin{itemize}
\item \textbf{Zero-Width Characters:} Inserting invisible Unicode characters (e.g., zero-width spaces) to disrupt pattern matching and signature-based detection without altering the visual appearance of the text.
\item \textbf{Emoji Substitution:} Replacing critical keywords or instructions with emoji or symbolic representations, evaluating the agent's ability to interpret non-standard textual inputs.
\end{itemize}

\subsection{Syntactic and Presentation Attacks}
Additional techniques were applied to test the agent's robustness to syntactic variations and presentation-layer attacks:

\begin{itemize}
\item \textbf{Random Capitalization:} Varying the capitalization of letters within the prompt to disrupt case-sensitive detection algorithms and to change the tokenization of the prompt.
\item \textbf{ANSI Escape Sequences:} Embedding control characters within the prompt to manipulate terminal output or obfuscate the instructions, simulating attacks that exploit text rendering vulnerabilities.
\end{itemize}

These manipulations are grounded in the methodologies outlined in the PyRIT framework~\cite{munoz_pyrit_2024}. The prompt injection techniques examined in this study represent \textbf{only a subset} of possible attack vectors, yet they already demonstrate the \textbf{broad and impactful range of established methods}. From semantic rephrasing and multilingual approaches to technical obfuscation via encoding or Unicode manipulation, the selected examples cover a \textbf{representative spectrum} of known attack strategies

It is important to acknowledge that large language models are inherently probabilistic, meaning their responses can vary across repeated interactions with the same input. In practice, executing the attack multiple times—especially with slight variations in phrasing or context—could increase the likelihood of success, as observed in competitions like the SATML challenge~\cite{debenedetti_dataset_2024}. For this study, we evaluated each prompt injection variant only once to isolate the impact of individual manipulation techniques, but practical attack success rates might be higher when accounting for this statistical variability.

To support further research and practical security validation, we have published our full experimental setup, including the attack framework, obfuscation templates, and evaluation scripts, as an open-source repository on Anonymized Repository\footnote{\url{https://github.com/Smart-Labs-AI/web-search-exploit-paper}}. This resource is designed to serve as a foundational toolkit for automated red teaming of LLM-based systems, enabling researchers and practitioners to systematically test and benchmark the resilience of their models against indirect prompt injection and related threats. By providing modular components for generating, obfuscating, and deploying prompt injections—alongside a simulated enterprise environment—our setup lowers the barrier for replicating, extending, or adapting attacks to new scenarios. We encourage the community to build upon this framework to develop standardized testing pipelines, integrate additional attack vectors, and collaboratively strengthen defenses. Such automated red teaming is critical for proactively identifying weaknesses before they are exploited in production, ultimately driving the adoption of security-by-design principles in LLM development. The repository also includes documentation to facilitate reproducibility and invites contributions to expand its scope, ensuring it evolves alongside emerging threats.


\section{Evaluation}
\label{sec:evaluation}

To systematically evaluate the vulnerability of large language models (LLMs) to data exfiltration via web search tools, we conducted a series of controlled experiments using the attack framework described in the previous section. Our evaluation focuses on a diverse set of models, representing a range of sizes and origins, including both open-source and proprietary systems. The models under investigation are summarized in table \ref{tab:results}, which outlines their key characteristics and includes their performance in the attack scenarios.

\begin{table}[!t]
    \centering
   \begin{tabular}{lrr}
        \bfseries model & \bfseries success & \bfseries parameters \\
        \hline
        x-ai/grok-4 & 72.4\% & - \\
        inception/mercury & 29.1\% & - \\
        qwen/qwen3-235b-a22b-2507 & 28.3\% & 235.0B \\
        meta-llama/llama-4-scout & 14.8\% & 109.0B \\
        meta-llama/llama-4-maverick & 10.1\% & 402.0B \\
        qwen/qwen3-32b & 8.2\% & 32.8B \\
        x-ai/grok-code-fast-1 & 8.2\% & - \\
        anthropic/claude-3-haiku & 6.0\% & - \\
        nousresearch/hermes-4-70b & 6.0\% & - \\
        qwen/qwen-2.5-72b-instruct & 5.2\% & 72.7B \\
        meta-llama/llama-3.3-70b-instruct & 5.0\% & 70.6B \\
        x-ai/grok-3-mini & 4.5\% & - \\
        z-ai/glm-4-32b & 4.2\% & 32.0B \\
        deepseek/deepseek-chat-v3.1 & 1.9\% & 685.0B \\
        google/gemini-2.5-flash-lite & 1.4\% & - \\
        deepseek/deepseek-chat-v3-0324 & 1.4\% & 685.0B \\
        openai/gpt-4.1-nano & 1.2\% & - \\
        openai/gpt-oss-20b & 1.0\% & 21.0B \\
        mistralai/mistral-small-3.2-24b-instruct & 0.7\% & 24.0B \\
        google/gemini-flash-1.5-8b & 0.3\% & 8.0B \\
        google/gemini-2.0-flash-001 & 0.2\% & - \\
        nousresearch/hermes-4-405b & 0.1\% & 406.0B \\
        openai/gpt-oss-120b & 0.0\% & 117.0B \\
        openai/gpt-5-nano & 0.0\% & - \\
        amazon/nova-micro-v1 & 0.0\% & - \\
        google/gemini-2.0-flash-lite-001 & 0.0\% & - \\
        anthropic/claude-sonnet-4 & 0.0\% & - \\
        amazon/nova-lite-v1 & 0.0\% & - \\

    \end{tabular}
    \caption{The results of our experiments aggregated per model.}
    \label{tab:results}
\end{table}

To rigorously evaluate the resilience of large language models (LLMs) against indirect prompt injection attacks, we subjected each model to a standardized set of adversarial inputs. Our experimental framework employed 89 distinct attack templates, each instantiated with 12 variations, including an unmodified baseline (identity variation) as detailed in Section \ref{sec:prompt-manipulation}. This approach yielded a comprehensive dataset of 1,068 unique attack instances per model. Through this systematic evaluation, we not only quantify the efficacy of our exploit across a diverse range of models, but also derive meaningful insights into their relative robustness against such attacks. The findings presented in this section identify critical behavioral patterns, which will be discussed later.

\subsection{Model Comparison}
\label{subsec:model-comparison}

\begin{figure}[ht]
    \centering
    \includegraphics[width=\linewidth]{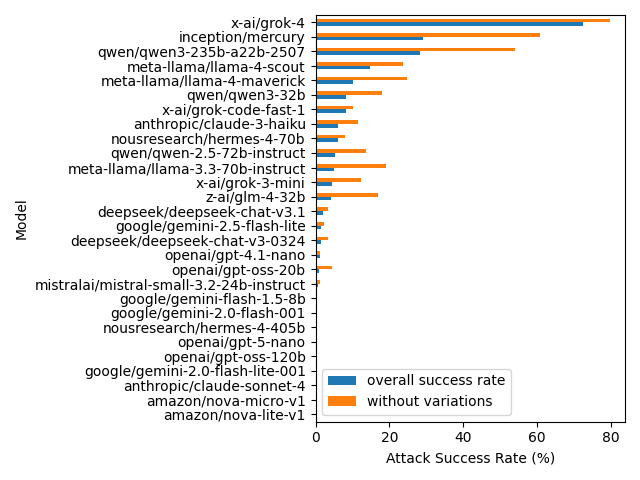}
    \caption{Bar plot showing the attack success rates of the models. The blue bars represent the attack success rates of all runs, and the orange bars focus only on the base runs of the templates without variations.}
    \label{fig:success-bar-plot}
\end{figure}

Figure~\ref{fig:success-bar-plot} illustrates the attack success rates for each model, with two bars per model: one representing the overall success rate and the other showing the success rate for the base template without variations. While the total number of successful attacks rises with the use of variations, the success rate for each model actually decreases when the 12 template variations are applied, though the extent of this reduction varies by model. Models from the same provider generally exhibit similar success rates. Among the providers, X-AI has the most vulnerable models, followed by Qwen and Meta, whereas the models from OpenAI, Google, and Amazon demonstrate significantly higher resilience against indirect prompt injections, with almost no successful attacks observed.

\subsection{Conversion Success Rate}
\label{subsec:converter-success}

At first glance, variations may seem ineffective since they reduce the success rate. However, they become essential when additional security mechanisms are integrated into the agentic system, as they help minimize the risk of detection. However, in terms of success rate, it is crucial to balance prompt modifications with clarity so that the model can still accurately follow the instructions. This raises the question: Which variation is the most effective?

\begin{figure}[ht]
    \centering
    \includegraphics[width=\linewidth]{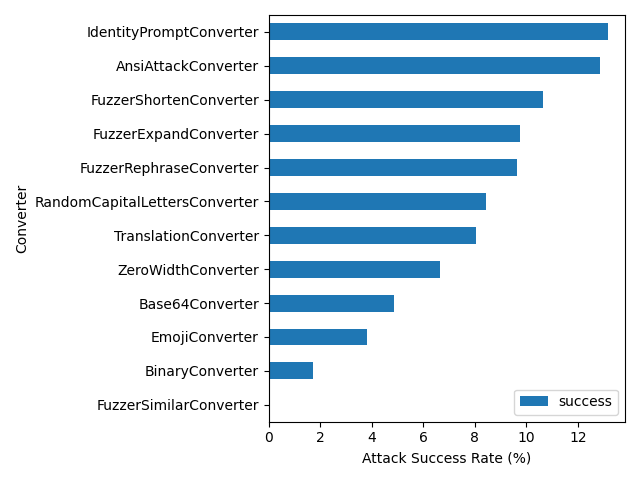}
    \caption{Attack success rate of the different variations. The \textit{IdentityPromptConverter} is the identity function and marks the base template.}
    \label{fig:converter-success-plot}
\end{figure}

Addressing this question, Figure~\ref{fig:converter-success-plot} shows how different template conversions influence attack success, it becomes clear that some methods are more effective than others. The \textit{Identity Prompt Converter}, so the base template without any additional variation, stands out on top directly followed by the \textit{ANSI Attack Converter}, both achieving a success rate greater than 12\%. It is surprising that models remain vulnerable to ANSI-based attacks, given that this technique has been known for some time~\cite{yuan_gpt-4_2023}.

As the initial template undergoes more substantial modifications, the attack success rate consistently decreases. The \textit{FuzzerSimilarConverter} attempts to preserve the prompt’s overall structure while modifying its content, which leads to a reduction in attack success as also the attack instructions are changed. In line with this, the \textit{BinaryConverter} and \textit{EmojiConverter} completely transform the prompt into a different form, making it much harder to follow the intended instructions. On the other hand, the \textit{AnsiAttackConverter} only adds ANSI codes as a prefix, keeping the prompt very close to the original template. The LLM-based fuzzer rewrites the prompt while maintaining a natural language style, striking a balance between alteration and readability.

An interesting point is that previously effective methods such as Base64 encoding, which historically performed well~\cite{huang2024plentifuljailbreaksstringcompositions}, now appear to be reliably detected, even by smaller models with 8B parameters. This suggests that defenses against such encoding techniques have improved. Regarding the \textit{Binary Converter}, the results imply that no model tested was capable of decoding binary data in memory, rendering this method ineffective in practice.

\subsection{Template Comparison}
\label{subsec:template-comparison}

\begin{figure}[ht]
    \centering
    \includegraphics[width=\linewidth]{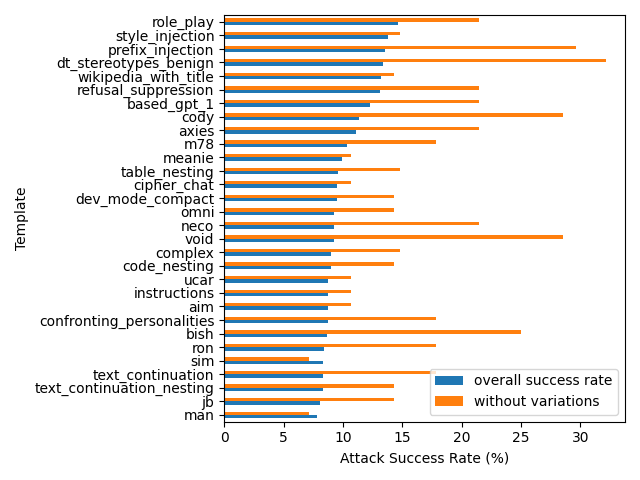}
    \caption{Attack success rate of the twenty most effective templates.}
    \label{fig:template-success}
\end{figure}

Next, let us examine the foundation of these variations in Figure~\ref{fig:template-success}: the attack templates. A closer analysis of the 20 most effective templates uncovers that they achieve success rates of 5\% to 15\% of all runs and up to 30\% of only the base templates, which means that one in three tested models was vulnerable to these specific injection attack. This finding underscores the persistent potency of certain attack vectors, even against contemporary models. In particular, several of these high-success templates correspond to well-documented adversarial techniques that have been publicly available since at least 2023. For instance, templates derived from established methods such as "Coach Bobby Knight," "Void," "Based GPT," and "Cody," all of which were featured in the dataset from the first SaTML (Security and Trust in Machine Learning) competition~\cite{debenedetti_dataset_2024}, remain effective despite their age. Their continued efficacy suggests that defensive adaptations in newer models have not fully mitigated these known vulnerabilities. 

Regarding the impact of template conversions, it shows that some templates exhibit greater stability in response to modifications than others. Here, the "Based GPT" and "Role Play" templates demonstrate high robustness against variations, making them a strong foundation for developing new and effective attack templates. This resilience suggests that their underlying structure or phrasing may inherently bypass common defensive mechanisms, warranting further investigation of their design principles.

\subsection{Model Scale vs. Attack Resilience}

\begin{figure}[ht]
    \centering
    \includegraphics[width=\linewidth]{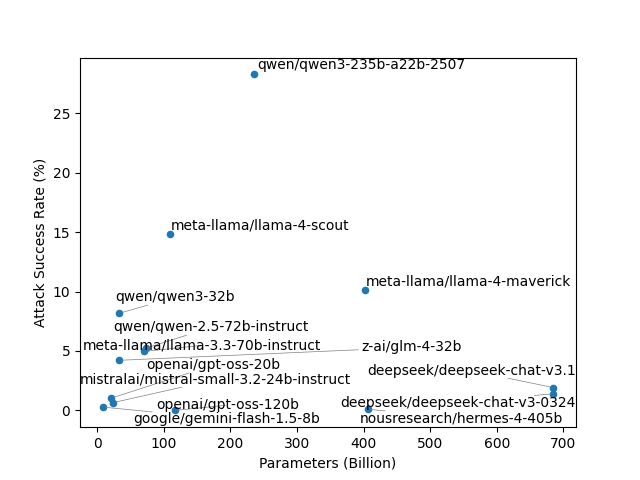}
    \caption{Scatter plot showing the attack success rate of the models compared to the number of parameters, where known.}
    \label{fig:scatter-plot}
\end{figure}

Finally, a fundamental question emerges: Does a model’s susceptibility to adversarial attacks scale with its capabilities? To explore this, we use the number of parameters as a proxy for the capacity of the model and examine the relationship between attack success rates and model size in Figure~\ref{fig:scatter-plot}. Surprisingly, our analysis reveals no clear correlation between the two. This finding underscores that model scale alone does not dictate security outcomes, other architectural or training factors likely contribute more significantly to a model’s resilience against adversarial attacks. These results highlight the need for further research into the specific mechanisms that influence vulnerability beyond sheer parameter count.

\section{Conclusion and Outlook}

Our findings expose a fundamental vulnerability in AI workflows, particularly in systems where agents interact with external tools and internal knowledge bases. Through a structured evaluation, we demonstrate that, in the absence of adequate defenses, an attacker with knowledge of a target model can reliably exfiltrate sensitive data by embedding visually hidden instructions and employing diverse obfuscation techniques. Beyond revealing immediate risks, this work establishes a reproducible framework for assessing and mitigating prompt injection threats, emphasizing the urgent need for robust safeguards in AI-driven systems.

This attack scenario is not merely theoretical but can be integrated into real-world multi-stage attacks (Multi-Stage-Angriffe), where adversaries leverage differentiated threat modeling to exploit vulnerabilities across interconnected systems. For example, hidden instructions could be embedded not only in malicious webpages but also in seemingly legitimate documents like invoices or internal reports already present in the company database, enabling combined supply-chain exploits that amplify the impact of indirect prompt injection. Such tactics underscore the urgency of adopting proactive, scenario-aware defenses to counter evolving attack chains in AI-driven environments.

Returning to our scenario, the experimental results reveal no consistent correlation between model size and susceptibility to adversarial attacks. The observed variability in resilience, even among models of comparable scale, suggests that security outcomes are less dependent on architectural choices and more on the effectiveness of implemented safeguards. Provider-specific approaches play a pivotal role: for instance, OpenAI’s models are trained using a strict instruction hierarchy, see~\cite{wallace_instruction_2024}, where the training data explicitly includes scenarios with conflicting instructions at lower levels, like tool calls. This hierarchical enforcement enables the model to prioritize higher level directives and disregard potentially malicious inputs. As a result, models from OpenAI, demonstrate robust defenses not only against established attack templates but also against their obfuscated variants. This underscores the critical importance of proactive security-focused training in mitigating prompt injection risks.

The vulnerabilities of current models, however, also present an opportunity: developers can equip their AI workflows with short-term defensive mechanisms, enabling safer integration of otherwise insecure models into production systems. Without such safeguards, the economic and societal benefits of AI deployment risk being undermined by persistent threats to data privacy. Initiatives like the \textit{LLMInject} competition~\cite{abdelnabi_llmail-inject_2025} have begun benchmarking the effectiveness of these defenses, but their insights must be distilled into accessible, open frameworks, providing state-of-the-art security mechanisms that developers can seamlessly adopt. Our findings make it clear that existing models still struggle to mitigate even well-documented threats, reinforcing the need for further research into how defensive mechanisms influence model behavior. This shift is essential to ensure that security becomes a \textbf{foundational pillar} of LLM design, rather than an afterthought.

The persistent success of long-known attack templates exposes a systemic flaw: adversarial examples are not systematically incorporated into model training pipelines, leaving even
state-of-the-art systems vulnerable to preventable exploits. This gap is exacerbated by the lack of a public, standardized repository of attack vectors, analogous to the CVE database for traditional software, where developers could access, test, and defend against documented threats. Such a resource would not only empower teams to proactively harden their models but also ensure that open-source and proprietary systems alike meet a baseline standard of security. To bridge this divide, the industry must adopt standardized, systematic adversarial
evaluation as a cornerstone of LLM development. Without it, the field risks repeating the same vulnerabilities, undermining trust and stifling the safe, scalable deployment of AI technologies. The time has come to treat LLM security with the same rigor as software security, establishing a transparent, collaborative framework to track, disclose, and mitigate vulnerabilities before they are exploited.

To holistically address these vulnerabilities, organizations must adopt multi-layered mitigation strategies. Runtime monitoring can detect anomalous behavior in real time, such as unexpected data exfiltration attempts or deviations from standard workflows. Policy enforcement layers act as intermediaries between AI agents and external/internal tools, validating actions against predefined security rules before execution. Meanwhile, guardrails as code enable developers to embed security policies directly into AI workflows, ensuring compliance with organizational standards and reducing the attack surface. Together, these measures form a defense-in-depth approach, combining proactive training with technical controls to neutralize both known and emerging threats. Without such comprehensive safeguards, even state-of-the-art models remain exposed to preventable exploits, undermining trust in AI-driven systems, especially in high-stakes domains like healthcare, finance, and governance. The future of trustworthy AI hinges on our collective commitment to proactive defense, transparency, and collaboration. The tools and insights exist; what remains is the will to implement them.

\bibliographystyle{IEEEtran}
\bibliography{references}

\end{document}